  \documentstyle[12pt]{article}
  
\catcode`@=11
\def\secteqno{\@addtoreset{equation}{section}%
\def\theequation{\thesection.\arabic{equation}}}
\catcode`@=12
\secteqno
\newcommand{\be}{\begin{equation}}
\newcommand{\ee}{\end{equation}}
\newcommand{\bea}{\begin{eqnarray}}
\newcommand{\eea}{\end{eqnarray}}
\newcommand{\bref}[1]{(\ref{#1})}
\newcommand{\ep}{\epsilon} 
\newcommand{\T}{\theta} 
   
\newcommand{\A}{\alpha} \newcommand{\B}{\beta}
\newcommand{\G}{\Gamma} \newcommand{\D}{\delta}
\newcommand{\k}{\kappa}         \newcommand{\lam}{\lambda}
           \newcommand{\s}{\sigma}
          
\newcommand{\z}{\zeta}
           \newcommand{\Th}{\Theta}

\def\pa{\partial}

\def\CL{{\cal L}}

\def\CF{{\cal F}}
\def\Tb{{\overline\theta}}
\def\Thb{{\overline\Theta}}

\def\tension{{{\rm T}}}
\newcommand{\nn}{\nonumber}

\def\Sigb{\overline\Sigma}

\def\ba{\overline}

\def\tp{\tilde p}
\def\t{\tilde}

\def\l{{\ell}}

\def\gg{{\tau_3}}
\def\CE{{{\cal C}_1}}
\def\SE{{{\cal S}_1}}
\def\fg{{~\{f\leftrightarrow g\}~}}

\newcommand{\sltp}{/ {\hskip-0.27cm{\tilde{p}}}}
\newcommand{\slPi}{/ {\hskip-0.27cm{\Pi}}}

\newcommand{\slB}{/ {\hskip-0.21cm{V}}}

\title{{\bf Canonical Formulation of IIB D-branes }}
\author{{\sc Kiyoshi Kamimura\  and
         Machiko Hatsuda$^{\dagger}$}\\
        \small{\it{Department of Physics, Toho University}}\\
        \small{\it{Funabashi, 274-8501 JAPAN}}\\
        \llap{$^\dagger$}
        \small{\it{ Department of Radiological Sciences,}}\\
        \small{\it{Ibaraki Prefectural University of Health Sciences}}\\
        \small{\it{Ami, Inashiki, Ibaraki, 300-0394 JAPAN}}\\
  \small{kamimura@ph.sci.toho-u.ac.jp,
         hatuda@ipu.ac.jp}\\
}
\date{}

\begin{document}

\maketitle

\thispagestyle{empty}

\begin{abstract}
We find Wess-Zumino actions for kappa invariant type IIB D-branes in 
explicit forms. A simple and compact expression is obtained by 
the use of spinor variables which are defined as power series 
of differential forms. Using the Wess-Zumino actions 
we develop the canonical formulation and find the complete set of 
the constraint equations for generic type IIB D-p-branes. 
The conserved global supersymmetry charges is determined 
and the algebra containing the central charges 
can be obtained explicitly.
\end{abstract}
\vskip 10mm

\vskip 5mm

\vfill
\vbox{
\hfill hep-th.9712068\null\par
\hfill December 1997\null\par
\hfill Revised in April 1998\null\par
\hfill TOHO-FP-9757\null\par
\hfill }\null

\clearpage
\section{Introduction}
\indent

The superstring theories\footnote{See \cite{GSW} for the review.} 
are promising approaches to the unified theory. 
In their non-perturvative approaches D-branes play important 
roles \cite{pol}\cite{Wit}.
D-branes, as well as superstrings, are characterized by 
the diffeomorphism, the local supersymmetries (kappa symmetry)
and the global super-Poincar$\acute{\rm e}$ invariance. 
Kappa invariant D-brane actions have been proposed in
\cite{Shkp}\cite{Shgf}\cite{Shdd}\cite{Cw}\cite{BT}.
The action of the D-branes consists of two parts, the Dirac-Born-Infeld (DBI) 
action and the Wess-Zumino (WZ) action. 
The DBI lagrangian is manifestly symmetric under SUSY transformation
while the WZ lagrangian is pseudo symmetric, i.e. invariant up to 
total derivative terms.
Under the kappa transformation only the sum of DBI and WZ actions 
remain invariant.  

In this paper we develop the canonical formulation of
the D-p-branes for generic $p$. 
It is the first step towards the quantum theories; 
in canonical, path integral, BRST quantizations and so on. 
It has been discussed the canonical quantization
in connection with the covariant gauge fixing 
for D-particle and D-string \cite{RK}\cite{MK}.

The defining equation of the WZ action for type IIB  D-p-brane 
in the flat background is, 
using the notation of \cite{Shgf}, as
\bea
d~L^{WZ}~&=&~~\tension~d\Tb~{\cal S}_B~\tau_1~d\T~e^\CF.
\label{wzcond}
\eea
The integrability is guaranteed by the Bianchi identity for the
Ramond/Ramond curvature $R=d\Tb{\cal S}_B\tau_1d\T$
and the WZ action is determined up to exact forms.
The WZ action $L^{WZ}_{p+1}$ of the IIB  D-p-brane 
is the $(p+1)$ form part of the WZ action $~L^{WZ}$.
\bref{wzcond} would be sufficient in most purposes;
the local and global supersymmetry invariances can be proven
using only the defining equation. 
In the case of super p-brane the algebras of covariant derivatives 
and SUSY were determined from the Ramond/Ramond curvature 
by using the covariance under supertranslations \cite{AIT}.
However 
the integrated form for $L^{WZ}$ is often desirable.
In the canonical quantization 
the form of the WZ action is required. 
For the general D-p-branes 
it had not been fully studied since the explicit form of the WZ action
was not available \cite{RK}\cite{LEE}.
In order to perform the canonical analysis for
the general D-p-branes, 
we obtain the WZ action in explicit form in this paper.

Integration of the differential equation \bref{wzcond} is not 
straightforward except for small p. 
Finding out appropriate variables are essential to perform this integration.
In this paper we introduce new spinor variables,
$\Th_C$ and $\Th_S$, which are power series of the differential forms.
After integrating \bref{wzcond}, the resultant WZ action is obtained as
\bea
L^{WZ}&=&\tension~(~\Thb_C~{\cal S}_B~+~\Thb_S~{\cal C}_B~)~\tau_1~e^\CF~d\T.
\label{solwz0}
\eea       
Using this we introduce canonical momenta and
derive the complete set of constraint equations for the IIB D-p-branes. 
We also derive the conserved SUSY charges 
and compute the SUSY algebra and the central charges 
for the IIB D-p-branes as functions of world volume variables.
As was discussed in \cite{AGIT}\cite{AIT} the SUSY algebras of extended 
objects can induce central extensions 
due to the pseudo-invariance of the WZ action
under the SUSY transformations.
The SUSY algebras exhibit their BPS properties and
discussions of the central charges are one of the 
important current interests \cite{BSW,Hull,sT}. 
Local and global supersymmetries of super D-1-brane are deeply related
as demonstrated in the reference \cite{MK}.
Such relation can be examined for D-p-branes with arbitrary p, 
once explicit forms of kappa constraints, SUSY charges and their algebras
are determined.
These analysis will give the correct treatments of fermionic degrees of
freedom leading to quantum D-brane theories. 
\medskip

In section 2 we introduce some crucial definitions and formulas
and integrate \bref{wzcond} to find the  WZ action $~L^{WZ}~$ \bref{solwz0}. 
In section 3 the constraint equations in the Hamiltonian formalism
are derived. The algebra of fermionic constraints is computed 
and the kappa symmetry of the system is assured in section 4.
In section 5 we show the quasi invariance
of the WZ lagrangian under the SUSY transformation.
The super symmetry charges are constructed by utilizing the
the surface term of the SUSY transformation.  
In section 6 the global supersymmetry algebra is 
computed and the expressions of the central charges are presented.
In the last section we give short summary and discussions. 
We add some detailed formulas in Appendix for convention. 
\vskip 6mm

\section{WZ action}
\indent

The actions of the super D-branes are described by the 10D target space
vector coordinates $X^m,~(m=0,...,9)$, Majorana-Weyl spinors $\T^\A_A,~
(A=1,2)$ and U(1) gauge field $A_\mu,~(\mu=0,..,p)$. In this paper
we consider general type IIB theory ( p=odd ) and the two Majorana-Weyl 
spinors (A=1,2) have the same chirality. 
The action is constructed from global SUSY invariant forms. They are 
$~d\T~$,
\bea
\Pi^\l &\equiv&dX^\l ~+~\Tb~\G^\l ~d\T~\equiv~d\s^\mu~\Pi_\mu^\l
\eea
and
\bea
\CF&\equiv&dA~+~\Tb~\gg\G_\l~d\T~(~\Pi^\l~-~\frac12~\Tb\G^\l~d\T~)~\equiv~
\frac12 d\s^\mu d\s^\nu~\CF_{\mu\nu}.
\label{defCF}
\eea
We follow the notations of \cite{Shgf} except their $\psi$ is denoted as
$\slPi=\Pi^m\G_m$. Remember the spinors anticommute with odd forms.

The DBI lagrangian density 
\bea
\CL^{DBI}&=&-~\tension~\sqrt{-det(G_{\mu\nu}+\CF_{\mu\nu})}.
\label{DBIL}
\eea
is invariant under global SUSY since the induced metric 
$G_{\mu\nu}=\Pi_{\mu}^m\Pi_{m\nu}$ and U(1) field strength $\CF$
are invariant.
On the other hand the SUSY transformation of the WZ lagrangian density 
is a total divergence. It is seen from \bref{wzcond} 
that $d L^{WZ}$ is given by a SUSY invariant form. 
In this section we solve the \bref{wzcond} to find the explicit form
of $L^{WZ}$.
\medskip

As preparations we give some useful definitions and formulas.
First one is ${\cal S}_B$ and ${\cal C}_B$ introduced in \cite{Shgf}
\bea
{\cal S}_B(\slPi)&=&\sum_{\l=0}~\gg^{\l}\frac{\slPi^{2\l+1}}{(2\l+1)!}~=~
\slPi~+~\gg~\frac{\slPi^3}{3!}~+~\frac{\slPi^5}{5!}~+~...~,
\\
{\cal C}_B(\slPi)&=&\sum_{\l=0}~\gg^{\l+1}\frac{\slPi^{2\l}}{(2\l)!}~~~~=~
\gg~+~\frac{\slPi^2}{2!}~+~\gg~\frac{\slPi^4}{4!}~+~...~.
\label{defS}
\eea

Second is $2\times 2$-matrix valued 10D vector defined for any
type IIB spinors $\psi$ and $\phi$
\bea
V^m_{\psi,\phi}&\equiv&(\ba\psi~\G^m~\phi)~+~\gg~(\ba\psi~\gg~\G^m~\phi).
\label{BBB}
\eea
Using this notation the cyclic identity holds as
\bea
\slB_{\psi,\phi}~\T~+~\slB_{\phi,\T}~\psi~+~\slB_{\T,\psi}~\phi~=~0
\label{cyc}
\eea
for three odd IIB spinors. It frequently appears an expression
\bea
V^m&\equiv&V^m_{\T,d\T}~=~(\ba\T~\G^m~d\T)~+~\gg~(\ba\T~\gg~\G^m~d\T).
\eea
The cyclic identity \bref{cyc} tells, for example, 
\be
d \slB~\T~+~2~\slB~d\T~=~0,~~~~~~~\rightarrow~~~~~~~~
d\slB~d\T~=~0,~~~~~~~~V\cdot dV~=~0.
\label{slbdt}
\ee
\medskip

Third is j-form IIB spinor $\Th_j$ defined by
\bea
\Th_{j}&\equiv&\slB~\Th_{j-1}~=...=~\slB^j~\Th_0,~~~~~~~~~~
\Th_{0}~\equiv~\T.
\eea
It has the $j+1$ parity . 
It holds an important formula 
\be
d~\Th_j~=~~\frac{2j+1}{2}~d\slB~\Th_{j-1}~,~~~~~~~~~(j=1,2,...)
\label{dTh}
\ee
which is proved by induction ( see Appendix \bref{dTh1}).
\medskip

The important quantities describing the D-branes 
are odd IIB spinor form $\Th_S$ and even one $\Th_C$ defined by
\bea
\Th_S&\equiv&\sum_{n=0}~\frac{1}{(4n+3)!!}~(-\gg)^{n+1}~\Th_{2n+1}~=~
-~\frac{1}{3!!}~\gg~\slB~\T~+~\frac{1}{7!!}~\slB^3~\T~-~...,
\nn\\
\Th_C&\equiv&\sum_{n=0}~\frac{1}{(4n+1)!!}~~(-\gg)^n~~\Th_{2n}~~~=~
\T~-~\frac{1}{5!!}~\gg~\slB^2~\T~+~... ~.
\label{defThCS}
\eea
Using \bref{dTh} they are shown to satisfy 
\bea
d~\Th_C~=~d\T~+~\frac12~d\slB~\Th_S,~~~~~~~~~~~
d~\Th_S~=~-~\frac12~\gg~d\slB~\Th_C.
\label{dthcs}
\eea
Here $d\T$ term appears from n=0 term.

The last is ${\cal S}_1$ and ${\cal C}_1$, which are respectively 
$T_B$ and $\rho_B$ of ref.\cite{Shgf},
\bea
{\cal S}_1~=~{\cal S}_B~\tau_1~e^\CF,~~~~~~~~~~~~
{\cal C}_1~=~{\cal C}_B~\tau_1~e^\CF.
\label{defs1c1}
\eea
They satisfy, using \bref{dchiCS1},
\bea
d~{\cal S}_1&=&\frac12~\gg~[~d~\slB~{\cal C}_1~+~{\cal C}_1~d~\slB~],
\nn\\
d~{\cal C}_1&=&\frac12~[~d~\slB~{\cal S}_1~-~{\cal S}_1~d~\slB~].
\label{dCS}
\eea
\medskip

Now we are ready to integrate \bref{wzcond}
\bea
d~L^{WZ}~&=&~+~\tension~d\Tb~{\cal S}_1~d\T
\label{WZcond}
\eea
to find the $L^{WZ}$.
We first write the right hand side of \bref{WZcond} as
\bea
d\Tb~{\cal S}_1~d\T&=&d~L_1~+~I_1,~~~~~~
L_1~=~\Tb~{\cal S}_1~d\T,~~~~~
I_1~=~\Tb~d~(~{\cal S}_1~d\T~).
\eea
Using \bref{dCS}, \bref{slbdt} and \bref{dTh}
\bea
I_1&=&\Tb~\frac12~d~\slB~\gg~{\cal C}_1~d\T~=~d~(~-\frac13~\Tb~\slB~)~
\gg~{\cal C}_1~d\T~\equiv~d~L_2~+~I_2
\eea
with
\bea
L_2~=~\frac13~\Thb_1~(-\gg)~{\cal C}_1~d\T~,~~~~~~~~~~
I_2~=~\frac13~\Thb_1~\gg~d~(~{\cal C}_1~d\T~).
\eea
Repeating this procedure  ( actually it terminates
at $(p+1)$-th step for the $p$ brane ) and summing up $L_j$'s 
we arrive at a compact expression
for the WZ action. It is, up to a total exterior
derivative term,  
\bea
L^{WZ}&=&\tension~(~\Thb_C~{\cal S}_1~+~\Thb_S~{\cal C}_1~)~d\T~,
\label{solwz}
\eea
where $\Th_S$ and $\Th_C$ are ones defined in \bref{defThCS}.
The WZ actions of the D-p-brane is the $(p+1)$ form part of \bref{solwz}.

It is straightforward to confirm \bref{WZcond} by
 using \bref{dthcs} and \bref{dCS}:
\bea
d~L^{WZ}&=&
\tension~[(d\Tb~-~\frac12\Thb_Sd\slB){\cal S}_1~-~
\Thb_C \{\frac12\gg(d\slB {\cal C}_1~+~{\cal C}_1 d\slB)\}
\nn\\
&+&(\frac12\Thb_C\gg d\slB){\cal C}_1~
+~\Thb_S\{\frac12(d\slB {\cal S}_1-{\cal S}_1d\slB)\}]d\T~
=~\tension~d\Tb {\cal S}_1 d\T~,
\eea
where only $d\Tb$ term in $d\Thb_C$ survives to give \bref{WZcond}.
\vskip 6mm


\section{Hamiltonian constraints}
\indent

In this section we develop the canonical formalism and
find the constraint equations for the IIB D-branes.

The lagrangian density of the system is 
\be
\CL~=~\CL^{DBI}~+~\CL^{WZ}
\ee
\be
\CL^{DBI}~=~-~\tension~\sqrt{-det(G_{\mu\nu}~+~\CF_{\mu\nu})}~\equiv~
-~\tension~\sqrt{-G_F}.
\ee
The WZ action has been determined in \bref{solwz}
\bea
\CL^{WZ}~=~[L^{WZ}]_{p+1},~~~~~
L^{WZ}&=&\tension~(~\Thb_C~{\cal S}_1~+~\Thb_S~{\cal C}_1~)~d\T~.
\label{solwz1}
\eea
where $[~~]_{p+1}$ means {p+1}-form coefficient ( coefficient of
$d\s^0...d\s^{p}$ in the p+1 form term ).
Let $a,b$ run from 1 to p.
We regard $\CL$ as a function of $\Pi_0,~\CF_{0a},~\dot\T$
and others,
\be
\CL~=~\CL^{DBI}(\Pi_0,~\CF_{0a},~\dot\T,...)~+~
\CL^{WZ}(\Pi_0,~\CF_{0a},~\dot\T,...).
\ee

We first define the canonical momenta $p_m,~\z,~E^\mu$
conjugate to $X^m$, $\T$ and
$A_\mu$. They are 
\bea
\left\{
\begin{array}{ccl}
p_m&\equiv&\frac{\pa\CL}{\pa \dot x^m}~
=~\frac{\pa\CL^{DBI}}{\pa \Pi^m_0}~+~\frac{\pa\CL^{WZ}}{\pa \Pi^m_0}~+~
\frac{\pa\CL}{\pa \CF_{0a}}~({\Tb~\gg~\G_m~\pa_a\T}),
\label{defp}
\\
\\
\z&\equiv&\frac{\pa^r\CL}{\pa \dot \T}~
=~(\frac{\pa\CL^{WZ}}{\pa \dot \T})~+~
(~\frac{\pa\CL}{\pa \Pi^m_0}~+~\frac{\pa\CL}{\pa \CF_{0a}}~
\frac{\pa \CF_{0a}}{\pa \Pi^m_0}~)~(-\Tb~\G^m)
\\
\\
&&+~\frac{\pa\CL}{\pa \CF_{0a}}~[-~\Tb~\gg\G_\l~
(~\Pi_{a}^\l~ +~ \frac12~\Tb\G^\l \pa_{a}\T~)~
+~ \frac12~\Tb\G^\l(\Tb~\gg\G_\l~\pa_{a}\T)],~
\label{defz}
\\
\\
E^a&\equiv&\frac{\pa\CL}{\pa \dot A_a}~=~\frac{\pa\CL}{\pa \CF_{0a}}~
=~\frac{\pa\CL^{DBI}}{\pa \CF_{0a}}~+~\frac{\pa\CL^{WZ}}{\pa \CF_{0a}}
\end{array}\right.
\label{defE}
\eea
and 
\bea
E^0&\equiv&\frac{\pa\CL}{\pa \dot A_0}~=~0.
\label{defE0}
\eea
The last one is U(1) constraint.

We show how the expressions for ${\pa\CL^{WZ}}/{\pa \Pi_{0}},~
{\pa\CL^{WZ}}/{\pa \CF_{0a}},~{\pa\CL^{WZ}}/{\pa \dot\T}~$ are
computed from $L^{WZ}$ without decomposing the differential forms.
Using the definition \bref{defCF} and \bref{defs1c1}, 
${\pa\CL^{WZ}}/{\pa \CF_{0a}}$ is  $p+1$ form coefficient of
\bea
\frac{\pa L^{WZ}}{\pa \CF_{0a}}&=&[d\s^0d\s^a~L^{WZ}~].
\eea
 Since explicit $\Pi$ appears only in ${\cal S}_1$ and ${\cal C}_1$ in  
$L^{WZ}$,
${\pa\CL^{WZ}}/{\pa\Pi_0^m}$ is (p+1) form coefficient of 
\bea
\frac{\pa L^{WZ}}{\pa\Pi_0^m}&=&-~T~ d\s^0[~
\Thb_C~(\gg~{\cal C}_1\G_m~-~\gg~{\cal S}_1~\Pi_m)~
\nn\\
&&~~~~~~~~~~+~\Thb_S~(~{\cal S}_1~\G_m~  -~\gg~{\cal C}_1~\Pi_m)~]~d\T,
\eea
where we have used \bref{DelSBCB}.
${\pa\CL^{WZ}}/{\pa \dot\T}$ is obtained from $L^{WZ}$ by 
using $\Delta\Thb_C$ and $\Delta\Thb_S$ in \bref{DelTC} and \bref{DelTS}.
Multiplying an odd spinor $f~$, 
$({\pa\CL^{WZ}}/{\pa \dot\T})f$ is (p+1) form coefficient of
\bea
\frac{\pa L^{WZ}}{\pa \dot\T}~f&=&T~d\s^0~[~(~\Thb_C~{\cal S}_1~+~
\Thb_S~{\cal C}_1~)~f~
\nn\\
\nn\\
&&~~~~~~~+~(~\Sigb^1_C[f]~{\cal S}_1~+~\Sigb^1_S[f]~{\cal C}_1~)~d\T~],
\label{WG}
\eea
where 
\bea
\Delta\T~=~0,~~~~\Delta V~=~V_{\T,d\s^0 f}~=~-~d\s^0~V_{\T,f}
\eea
in \bref{DelTC} and \bref{DelTS} and
the explicit forms of $\Sigb_C^1[f]$ and $\Sigb_S^1[f]$ are given 
in the Appendix \bref{tSig1SD}.
\medskip


In the Hamiltonian constraints the momenta appear in the following 
combinations, $\tilde E^a$ and $\tp_m$,
\bea
\left\{
\begin{array}{ccl}
\tp_m&\equiv&p_m~-~\frac{\pa\CL^{WZ}}{\pa \Pi^m_0}~-~
E^a~({\Tb~\gg~\G_m~\pa_a\T})
\\
\\
\tilde E^a&\equiv&E^a~-~\frac{\pa\CL^{WZ}}{\pa \CF_{0a}}
\end{array}\right..
\eea
Since $\CL^{WZ}(\Pi_0,~\CF_{0a},~\dot\T,...)$ is linear in the velocities
by construction, they are functions of canonical variables. 
\bref{defp} tells that they are contributions 
from the DBI action to the momenta and are 
\bea
\left\{
\begin{array}{ccl}
\tp_m&=&\frac{\pa\CL^{DBI}}{\pa \Pi^m_0}~=~
-~\frac{\tension}{2}~\sqrt{-G_F}~(~G_F^{\mu 0}~+~G_F^{0 \mu}~)~\Pi_{\mu m}
\\
\\
\tilde E^a&=&\frac{\pa\CL^{DBI}}{\pa \CF_{0a}}~=~
-~\frac{\tension}{2}~\sqrt{-G_F}~(~G_F^{a 0}~-~G_F^{0 a}~)
\end{array}\right..
\eea
It follows p+1 bosonic constraints
\bea
T_a&\equiv&\tp~\Pi_a~+~\tilde E^b~\CF_{ab}~=~0,~~~~~~~~~~~(a=1,2,...p)
\label{defTa}
\\
\nn\\
H&\equiv&\frac12~[~\tp^2~+~\tilde E^a~G_{ab}~\tilde E^b~+~
\tension^2 {\bf G_F}~]~=~0,
\label{defH}\eea
where  
\be
{\bf G_F}~=~\det (G+\CF)_{ab}.
\ee
\medskip
 The fermionic constraint follows from the definition of 
the momentum $\z$ in \bref{defz}, 
\bea
F&\equiv&\z~+~p_\l~(\Tb~\G^\l)~+~E^a~[~\Tb~\gg\G_\l~
(~\Pi_{a}^\l~ +~ \frac12~\Tb\G^\l \pa_{a}\T~)~
\nn\\
&&
-~ \frac12~\Tb\G^\l(\Tb~\gg\G_\l~\pa_{a}\T)]~-~
(\frac{\pa\CL^{WZ}}{\pa \dot \T})~=~0.
\label{defF}
\eea
The primary constraints are \bref{defTa}, \bref{defH}, \bref{defF}
and \bref{defE0}.  
The primitive forms of these constraint equations have been presented in 
\cite{RK} in connection with the discussions of the covariant 
quantization.
\medskip

The generalized Hamiltonian ${\cal H}~=~p\dot q-L$ is calculated only using 
$(p-{\pa L}/{\pa \dot q})^2\equiv 0$ \cite{kk} as
\bea
{\cal H}&=&\int d\s^p[~
\lam^0~H~+~\lam^a~T_a~+~F~\dot\T~+~E^0~\dot A_0~-~A_0~\pa_aE^a~],
\label{hamil0}\eea
where
\be
\lam^0~=~\frac{-1}{\tension\sqrt{-G_F}G^{00}_F},~~~~~~~~~~~~
\lam^a~=~\frac{G^{a0}_F+G^{0a}_F}{-2G^{00}_F}.
\ee
The consistency condition that the primary U(1) constraint \bref{defE0} 
is preserved in time requires secondary Gauss law constraint   
\bea
\pa_a~E^a~=~0.
\label{gauss}
\eea
The consistency of other constraints requires the form of 
$\dot\T$ in the Hamiltonian as
\bea
\dot\T~=~(\lam^0~\t\tau^a~+~\lam^a)~\pa_a\T~+~\Xi~\chi,
\label{dottheta}
\eea
where
\bea
\t\tau^a~=~\t E^a~\gg~+~{\tension}~\gg~[d\s^a~{\cal C}_1]_{\bf p}
\label{taua}
\eea
and $~\Xi~$ is a nilpotent matrix appearing in the  Poisson bracket of $F$'s
(see next section) 
\bea
\Xi&\equiv&\sltp~+~\slPi_a~\t E^a~\gg~+~\tension~[{\cal S}_1]_{\bf p},
\label{defchi}
\eea
and $\chi$ is an arbitrary spinor coefficient. 
Here $[....]_{\bf p}$ is the spatial p form coefficient 
( coefficient of $d\s^1...d\s^p$ ) of the expression $[....]$.
\bref{dottheta} is equivalent to the field equation of
$\T$. Using it the Hamiltonian \bref{hamil0} is written as
\bea
{\cal H}&=&\int d\s^p[~\lam^0~\t H~+~\lam^a~\t T_a~+~\t F~\chi~+~
E^0~\dot A_0~-~A_0~\pa_aE^a~],
\label{hamil1}\eea
where
\bea
\left\{
\begin{array}{ccl}
\t H&=&H~+~F~\t\tau^a~\pa_a\T,
\\
\\
\t T_a&=&T_a~+~F~\pa_a\T~=~p~\pa_ax~+~\z~\pa_a\T~+~E^b~F_{ab},
\\
\\
\t F&=&F~\Xi.
\end{array}\right.
\eea

The fact that all constraints are preserved in time for arbitrary 
undetermined multipliers $\lam^\mu,~\chi,~A_0$ and $\dot A_0$
means that the constraints $\t H,~\t T_a,~\t F,~E^0$ and $\pa_aE^a$
are the first class constraints. 
$\t H$ and $\t T_a$ are the first class constraints generating
the $p+1$ dimensional diffeomorphism.
$E^0$ and $\pa_aE^a$ are the standard U(1) generators.
As will be shown explicitly in the next section 
the fermionic constraint $F$ is a mixture of first and second class 
constraints as in the case of the superstring theories.
The half components $\t F$ are the first class constraints
generating the kappa transformations and other half are the second class ones. 
In the next section we find the Poisson bracket of $F$'s and 
examine the kappa symmetry using the explicit form of 
the WZ actions \bref{solwz1}.
\vskip 6mm


\section{Algebra of $F$'s}
\indent

To find the (graded) Poisson bracket algebra of the fermionic constraints  
we define $F[f]$ by multiplying a well behaved odd function $f$
\bea
F[f]&\equiv&\int~d\s~F(\s)~f(\s)
\nn\\
&=&\int~d\s~[\z f~+~p_m~(\Tb~\G^m~f)~
+~E^a~\{~\Tb~\gg\G_\l~f~
(~\Pi_{a}^\l~ +~ \frac{1}2~\Tb\G^\l \pa_{a}\T~)~
\nn\\
&&-~\frac12~\Tb\G^\l~f~(\Tb~\gg\G_\l~\pa_{a}\T)\}]~+~
\int~d\s~[-~\frac{\pa L^{WZ}}{\pa \dot\T}f~]
\nn\\
&\equiv&F^{(1)}[f]~+~F^{(2)}[f],
\eea
where $\frac{\pa L^{WZ}}{\pa \dot\T}f$ is given in \bref{WG}
and is a function of only coordinate variables. 
\medskip
$F^{(1)}$ is the WZ action independent part and satisfies
\bea
\{~F^{(1)}[f],~F^{(1)}[g]~\}&=&\int d\s~[~
-2~(\ba f\G^m~g)(p_m-E^b\Tb\gg\G_m\pa_b\T)~-~2~
(\ba f\gg\G_mg)(E^b~\Pi^m_b)
\nn\\
&+&\frac12(\pa_aE^a)\{(\Tb\G f)(\Tb\gg\G g)~-~(\Tb\G g)(\Tb\gg\G f)\}~].
\label{f1f1}
\eea
Here the last term is the Gauss law constraint term.

The WZ action dependent part $F^{(2)}$ does not contain momentum and 
the Poisson bracket with itself
vanishes,
\bea
\{~F^{(2)}[f],~F^{(2)}[g]~\}~=~0
\eea
while
\bea
\{~F^{(1)}[f],~F^{(2)}[g]~\}~+~\{~F^{(2)}[f],~F^{(1)}[g]~\}&=&
\nn\\
\int d\s~[~f^t\frac{\pa^\l}{\pa\T}(\frac{\pa L^{WZ}}{\pa \dot\T}g)~+~
\pa_a f^t\frac{\pa^\l}{\pa\pa_a\T}(\frac{\pa L^{WZ}}{\pa \dot\T}g)
&-&2~(\ba f\G^m\pa_a\T)~\frac{\pa^\l}
{\pa\Pi_a^m}(\frac{\pa L^{WZ}}{\pa \dot\T}g)~
\nn\\
-2~(\ba f\gg\G_\l\pa_a\T~\Pi^\l_b)~
\frac{\pa^\l}{\pa\CF_{ab}}(\frac{\pa L^{WZ}}{\pa \dot\T}g)~]&-&\fg,
\eea
where $\fg$ is terms in which $f$ and $g$ are interchanged.
The sum of these terms gives the Poisson bracket of $F$'s,
\bea
\{F[f],~F[g]\}~=~\int d\s~[~-2~(~\ba f~\Xi~g~)~+~
\frac12(\pa_aE^a)\{(\Tb\G f)(\Tb\gg\G g)-\fg\}].
\eea
The second term is the Gauss law constraint term and $\Xi$ is given
by 
\bea
\Xi&\equiv&\sltp~+~\gg~\slPi_a~\t E^a~+~\tension~\Xi^{(1)},~~~~~~~~~~~
\Xi^{(1)}~\equiv~[{\cal S}_1]_{\bf p}.
\eea

We can show that $\Xi$ is nilpotent on the bosonic primary constraints
\bref{defTa} and \bref{defH}.
It is reflecting the fact that
the half components of the fermionic constraints are the
first class while the remaining half are the second class
constraints. By squaring 
\bea
(\Xi)^2&=&(~\sltp~+~\gg~\slPi_a~\t E^a~+~\tension~\Xi^{(1)}~)^2.
\eea
The squares of each terms are
\bea
\sltp^2&=&\tp^2,~~~~~~~
(~\gg~\slPi_a~\t E^a~)^2~=~\t E^a~\t E^b~G_{ab},~~~~~~~
(\Xi^{(1)})^2~=~{\bf G}_F.
\eea
The last equality is essentially shown in the appendix A of \cite{Shgf}.
The sum of these three terms vanish on the bosonic primary constraint
$H$ in \bref{defH}.
The cross terms are 
\bea
\sltp~(\gg~\slPi_a~\t E^a)~+~(\gg~\slPi_a~\t E^a)~\sltp&=&
\gg~\t E^a~(~2~\tp~\Pi_a~)~=~2~\gg~\t E^a~T_a
\nn\\
\sltp~\Xi^{(1)}~+~\Xi^{(1)}~\sltp &=&-~2~[(\tp~\Pi)~{\cal C}_1~\tau_3]_{\bf p},
\nn\\
\gg~\slPi_a~\t E^a~\Xi^{(1)}~+~\Xi^{(1)}~\gg~\slPi_a~\t E^a&=&-~
2~[(\t E^b~\CF_{ab}~d\s^a)~{\cal C}_1~\tau_3]_{\bf p}.
\label{ct3}
\eea
The first one vanishes on the constraint \bref{defTa}.
The sum of second and third terms also vanish on the constraint $T_a$.
In summary
\bea
(\Xi)^2~=~2~H~+~2~\t\tau^a~T_a~,
\eea
where
$\t\tau^a$ is defined in \bref{taua}.
It shows the $\Xi$ is used to construct the first class constraints,
\be
\t F~=~F~\Xi,
\ee
which are the generators of the kappa transformations as in the case of 
the superstring theories.

\vskip 6mm

\section{Global SUSY transformations}
\indent

The global supersymmetry transformations of $X$ and $\T$ are
\bea
\D_\ep~\T&=&\ep,~~~~~~~~\D_\ep~X^m~=~\ba\ep~\G^m~\T.
\eea
For the U(1) potential it is defined so that its field strength
$\CF$ is SUSY invariant \cite{Shgf} as
\bea
\D_\ep~A&=&\ba\ep~\gg~\G~\T~d~X~-~\frac16~\ba\ep~\gg~\slB~\T.
\eea
In the canonical formalism the supersymmetry generator is
\bea
Q~\ep&=&\int d\s^p~(p_m~\D_\ep~X^m~+~\z~\D_\ep~\T~+~E^a~\D_\ep~A_a)~-~
\int d\s^p~F^0_\ep~
\nn\\
&\equiv&Q^1_\ep~+~Q^2_\ep.
\eea
$Q^1$ is  the WZ action independent part and is
\bea
Q^1_\ep~=~\int d\s^p~[~\z~\ep&-&\Tb~\G^m~\ep~(~p_m~-~
\frac16~\Tb~\gg~\G_m~\pa_a\T~E^a~)
\nn\\
&-&\Tb~\gg~\G^m~\ep~E^a~(~\pa_a~X_m~-~
\frac16~\Tb~\G_m~\pa_a\T~)].
\label{q1charge}
\eea
$F^0_\ep$ is determined from the surface term of the SUSY variation of 
the lagrangian,
\bea
\D_\ep~L^{WZ}~\equiv~d~(~\tension~[q^2_\ep]~),~~~~
{\rm and}~~~~F^0_\ep~=~\tension~[q^2_\ep]_{\bf p}.
\eea
where $[q^2_\ep]_{\bf p}$ is spatial p-form coefficient
of $[q^2_\ep]$ and
\bea 
Q^2_\ep~=~-~\tension~\int d\s^p~[q^2_\ep]_{\bf p}.
\label{q2charge}
\eea

In the WZ action \bref{solwz} 
$~{\cal S}_1,~{\cal C}_1$ and $d\T$ are SUSY invariant only $\Thb_C$ and 
$\Thb_S$
are varied
\bea
\D_\ep~L^{WZ}&=&\tension~(~\D_\ep\Thb_C~{\cal S}_1~+~
\D_\ep\Thb_S~{\cal C}_1~)~d\T~
=~\ba \Psi_S~d\T,
\label{delqlwz1}
\eea
where we define
\bea
\ba \Psi_S&\equiv&\tension~(~\D_\ep\Thb_C~{\cal S}_1+
\D_\ep\Thb_S~{\cal C}_1~),~~~~~
\ba \Psi_C~\equiv~\tension~(~-~\D_\ep\Thb_C~{\cal C}_1~\gg+
\D_\ep\Thb_S~{\cal S}_1~).~~~
\eea
Using \bref{dthcs} and \bref{dCS} they satisfy 
\bea
d~\ba \Psi_S&=&-~\frac12~\ba \Psi_C~d\slB,~~~~~~~~~~
d~\ba \Psi_C~=~-~\frac12~\ba \Psi_S~\gg~d\slB.
\label{dQ}
\eea
Writing \bref{delqlwz1} as
\bea
\D_\ep~L^{WZ}&=&d[~\ba \Psi_S~\T~]~-~(d\ba \Psi_S)~\T
\eea
and use \bref{dQ} and \bref{dTh}. As in the procedure leading to
\bref{solwz} we continue to arrive at  
\bea
\D~L^{WZ}&=&d~[~\ba \Psi_S~\Th_C~+~\ba \Psi_C~\gg~\Th_S~].
\label{SUSYLWZ}
\eea

Up to now we have only used the fact that $\D_\ep$ is even variation and 
the transformation is global, $~\D_\ep d\T~=~0~$.
The forms of SUSY variation $~\D_\ep\Thb_C~$ and $~\D_\ep\Thb_S~$ are
\bea
\D_\ep\Thb_C&=&\ba\ep~+~d~\Sigb_C^1[\ep]~+~\Sigb_C^2[\ep],~~~~~~~~~
\D_\ep\Thb_S~=d~\Sigb_S^1[\ep]~+~\Sigb_S^2[\ep],
\label{delThSC}
\eea
where $\Sigb_C^1[\ep]$ and $\Sigb_S^1[\ep]$ are given 
in the Appendix \bref{tSig1SD}
and $\Sigb_C^2[\ep]$ and $\Sigb_S^2[\ep]$ are satisfying
\bea
\Sigb^2_C[\ep]&=&
-~\frac{1}{2}~\Sigb^1_S[\ep]~d\slB~-~\Thb_S~\slB_{\ep,d\T},
\nn\\
\Sigb^2_S[\ep]&=&
(~\frac{1}{2}~\Sigb^1_C[\ep]~d\slB~-~\Thb_C~\slB_{\ep,d\T}~)~\gg.
\label{tSig2D}
\eea
Using them we can determine the surface term appearing in the SUSY 
variation of the WZ lagrangian,
\bea
\D_\ep~L^{WZ}~=~
\tension~d&[&\ba\ep~(~{\cal S}_1~\Th_C~-~{\cal C}_1~\Th_S~)~
+~(\Sigb_C^1[\ep]~{\cal S}_1~+~\Sigb_S^1[\ep]~{\cal C}_1)~d\T~
\nn\\
&-&\Thb_S~\slB_{\ep,d\T}~{\cal S}_1~\Th_C~+~
\Thb_S~\slB_{\ep,d\T}~{\cal C}_1~\Th_S~
\nn\\
&+&\Thb_C~\slB_{\ep,d\T}~{\cal S}_1~\Th_S~+~
\Thb_C~\slB_{\ep,d\T}~{\cal C}_1~\gg~\Th_C~]~\equiv~\tension~d~[q^2_\ep].
\label{SUSYWZ}
\eea
It gives $[q^2_\ep]$ and $Q^2_\ep$ is determined from \bref{q2charge}.
\vskip 6mm


\section{SUSY algebra and Central charges}
\indent

As was discussed in \cite{AGIT}\cite{AIT}
the SUSY algebras of extended objects
can induce central extensions 
by the pseudo-invariance of the WZ action
under the SUSY transformations.
We discuss it in the present case of D-p-branes. 

We have determined the supersymmetry charge $~Q=Q^1+Q^2~$ for general IIB
D-branes in \bref{q1charge} and \bref{q2charge} with $~q^2_\ep~$ given in
\bref{SUSYWZ}, up to an exact form, as
\bea
q^2_\ep&=&\ba\ep~(~{\cal S}_1~\Th_C~-~{\cal C}_1~\Th_S~)~
+~(\Sigb_C^1[\ep]~{\cal S}_1~+~\Sigb_S^1[\ep]~{\cal C}_1)~d\T~
\nn\\
&-&\Thb_S~\slB_{\ep,d\T}~{\cal S}_1~\Th_C~+~
\Thb_S~\slB_{\ep,d\T}~{\cal C}_1~\Th_S~
\nn\\
&+&\Thb_C~\slB_{\ep,d\T}~{\cal S}_1~\Th_S~+~
\Thb_C~\slB_{\ep,d\T}~{\cal C}_1~\gg~\Th_C.
\label{Q2}
\eea
In this section we will find the SUSY algebra and determine
the central charges in terms of the dynamical variables $X^m$, $\T$ and
$A_\mu$.

In the (graded) Poisson bracket of the supersymmetry charges,
$\{Q^1_\ep,~Q^1_{\ep'}\}~$
is independent of the WZ action and is common for all IIB D-branes, 
\bea
\{~Q^1_\ep,~Q^1_{\ep'}~\}&=&\int d\s^p~[~2~(\ba\ep~\G~\ep')~p~+~
2~(\ba\ep~\gg~\G~\ep')~\pa_a X~E^a
\nn\\
&-&\frac12(\pa_a~E^a)~(\Tb~\gg~\G~\ep~\Tb~\G~\ep'~-~
\Tb~\gg~\G~\ep'~\Tb~\G~\ep)~].
\eea
The first term is the total momentum term of the SUSY algebra. 
The last term is the Gauss law constraint term.
Since $Q^1$ generates SUSY transformations for coordinate variables 
and $Q^2$ contains only coordinates 
\bea
\{~Q^1_\ep,~Q^2_{\ep'}~\}&+&\{~Q^2_\ep,~Q^1_{\ep'}~\}~=~
-~\D_\ep~Q^2_{\ep'}~+~\D_{\ep'}~Q^2_{\ep}~\equiv~\D_{[\ep'}~Q^2_{\ep]},
\label{q2q2}
\eea 
\bea
\{~Q^2_\ep,~Q^2_{\ep'}~\}~=~0,
\eea
where our anti-symmetrization convention is 
$A_{[\ep}B_{\ep']}~=~A_{\ep}B_{\ep'}-A_{\ep'}B_{\ep}~$.
Thus the SUSY algebra becomes
\bea
\{~Q_\ep,~Q_{\ep'}~\}&=&\int d\s^p~[~2~(\ba\ep~\G~\ep')~p~+~
2~(\ba\ep~\gg~\G~\ep')~\pa_a X~E^a
\nn\\
&-&\frac12(\pa_a~E^a)~(\Tb~\gg~\G~\ep~\Tb~\G~\ep'~-~
\Tb~\gg~\G~\ep'~\Tb~\G~\ep)~]
\nn\\
&-&\tension~\int d\s^p~[~\D_{[\ep'}~q^2_{\ep]}~]_{\bf p}.
\label{QeQe}
\eea

We calculate the last term in the differential form 
$\D_{[\ep'}q^2_{\ep]}~$.
Since $~d\T,~\Pi,~\CF~$ are SUSY invariant, 
only the explicit $\T$'s are varied in $~q^2_{\ep}$. 
We use \bref{delThSC} and \bref{tSig2D} for $\D_\ep\Thb_C$ and $\D_\ep\Thb_S$
and
\bea
\D_{[\ep'}(\Sigb^1_C[\ep_{]}])~\equiv~d\Lambda^1_C~+~\Lambda^2_C,~~~~~~~~~
\D_{[\ep'}(\Sigb^1_S[\ep_{]}])~\equiv~d\Lambda^1_S~+~\Lambda^2_S~,
\label{delfSig}
\eea
where the explicit form of $\Lambda^1$'s are given in Appendix \bref{Lam1}.
$\Lambda^2$'s are  related to $\Lambda^1$'s  as
\bea
\Lambda^2_S&=&-~\frac12\Lambda^1_C\gg d\slB~-~\Thb_C\gg\slB_{\ep',\ep}~-~
\Sigb^1_C[{}_{[}\ep']\gg\slB_{\ep],d\T}
\nn\\
\Lambda^2_C&=&\frac12\Lambda^1_S d\slB~+~\Thb_S\slB_{\ep',\ep}~-~
\Sigb^1_S[{}_{[}\ep']\slB_{\ep],d\T}.
\label{lam2}
\eea
After some manipulation we obtain, up to an exact form,
\bea
\D_{[\ep'}~q^2_{\ep]}&=&
2~[~(\ba\ep~-~\Thb_S~\slB_{\ep,d\T}~)~\SE~(\ep'~+~\slB_{\ep',d\T}~\Th_S~)
\nn\\
&&+~(\ba\ep~-~\Thb_S~\slB_{\ep,d\T}~)~\CE~\gg~\slB_{\ep',d\T}~\Th_C~
+~\Thb_C~\slB_{\ep,d\T}~\CE~\gg~(\ep'~+~\slB_{\ep',d\T}~\Th_S~)
\nn\\
&&-~\Thb_C~\slB_{\ep,d\T}~\SE~\gg~\slB_{\ep',d\T}~\Th_C~].
\label{QQQ2}
\eea
\medskip

We will express the SUSY algebra in the standard form,
\bea
\{~Q_{\A A},~Q_{\B B}~\}&=&-~2~(C\G_m)_{\A\B}\D_{AB}~{\cal P}^m~
+~(C\G_M)_{\A\B}~(\tau_J)_{AB}~Z^M_J
\label{qqz}\eea
where the total momentum term is separated as usual and 
$(C\G_M\tau_J)$ is the complete basis of gamma matrices and tau 
matrices\cite{GSW}. $J$ runs from 0, 1, 2 and 3,
\be
tr~\frac{1}{2}(~\tau_J~\tau_L~)~=~\D_{JL},~~~~~~~~~~~(J,L=0,1,2,3).
\ee
$\G_M$'s are vectors $\G_ m$,  
totally anti-symmetric rank three tensors $\G_{ m_1 m_2 m_3}$ and 
totally anti-symmetric (anti-) self-dual rank five tensors 
$\G_{ m_1 m_2 m_3 m_4 m_5}$.
They are $10+120+252/2=16^2$  orthonormal basis for gamma matrices 
with the definite Weyl projection, 
(the projection operators are abbreviated throughout),
\be
tr~\frac{1}{2^4}((\G^N C^{-1})(C~\G_M))~=~\D_M^N
\ee
where $tr 1~=~2^4$ in the subspace by the Weyl projection.
We reverse the order of tensor indices for the $\G^M $. 
For example 
\bea
tr~\frac{1}{2^4}((~\G^{n_3n_2n_1}~C^{-1})(C~\G_{ m_1 m_2 m_3}))~=~
(\D_{ m_1}^{n_1} \D_{ m_2}^{n_2} \D_{ m_3}^{n_3}~+~ ...)
\eea
where ... means anti-symmetrized terms.
\medskip

Using \bref{QeQe} we obtain the SUSY algebra 
\bea
\{~Q_{\A A},~Q_{\B B}~\}&=&
-~2~(C\G_m)_{\A\B}~\D_{AB}~{\cal P}^m
\nn\\
&-&2~(\tau_3)_{AB}~(C\G_m)_{\A\B}~\int d\s^p~(~\pa_a X~E^a~)~
\nn\\
&+&\frac{1}{4}~(\tau_1)_{AB}(C\G_{m})_{\A\B}~
\int d\s^p(\pa_aE^a)~\Tb~\G^m~(i\tau_2)~\T
\nn\\
&-&\frac{1}{8}~(i\tau_2)_{AB}(C\G_{m_1m_2m_3})_{\A\B}~
\int d\s^p(\pa_aE^a)~
\Tb~\G^{m_3m_2m_1}~\tau_1~\T
\nn\\
&-&(\t\tau_J)_{AB}(C\G_{M})_{\A\B}\frac{2\tension}{2^5}(1+(-)^{\s_M+J})~
\int d\s^p~[U^M_J]_{\bf p}.
\label{qqal}
\eea
The first term is the total momentum term.
The second term is diagonal $\tau_3$ term giving $Z_3^m$,
\bea
Z_3^m~=~-~2~\int d\s^p~(~\pa_a X~E^a~).
\label{z3m}
\eea
Regarding the Gauss law constraint \bref{gauss} it is 
a topological charge.
The third and forth terms are the Gauss law constraint terms of \bref{QeQe}
and vanish on shell.
(The sum of ${m_1m_2m_3}$ is only over independent anti-symmetric 
indices.) 
The last term is the contribution from the WZ action, 
i.e. from  \bref{QQQ2} and depends on the
form of the WZ action explicitly.
Here $\t\tau_J=\tau_J~$ for $~J=1,2~$ and $\t\tau_J=0$ otherwise.
$\s_M=1$ for $M=m$ and $M={ m_1 m_2 m_3 m_4 m_5}$, and $\s_M=0$ for 
$M={ m_1 m_2 m_3}$. 
It shows it contributes only to off diagonal $\{ Q_1,Q_2\}$
and the vectors and 
the selfdual totally antisymmetric rank 5 tensors associate to $\tau_1$ and
the totally antisymmetric rank 3 tensors associate to $\tau_2$.  
It is consistent with the fact that \bref{qqz} is symmetric under simultaneous 
exchange of $(\A A)$ and $(\B B)$. 
The possible central charges are \bref{z3m} and
\bea
Z^m_1&=&-~\frac{\tension}{8}~\int d\s^p~[U^m_1]_{\bf p}~
-~\frac14~\int d\s^p(\pa_aE^a)~\Tb~\G^m\tau_1\gg\T~,
\nn\\
Z^{ m_3 m_2 m_1}_2&=&-~\frac{\tension}{8}~\int d\s^p~[U^
{ m_3 m_2 m_1}_2]_{\bf p}
-~\frac{1}{8}~\int d\s^p(\pa_aE^a)~
\Tb~\G^{ m_3 m_2 m_1}\tau_2\gg\T.
\nn\\
Z^{ m_5 m_4 m_3 m_2 m_1}_1&=&-~
\frac{\tension}{8}~\int d\s^p~[U^{ m_5 m_4 m_3 m_2 m_1}_1]_{\bf p}.
\eea
where $U$'s are given from \bref{QQQ2} as
\bea
U^M_J&=&\frac12~Tr~(~\t\tau_J~\G^M~{\cal S}_1~)
\nn\\
&+&2~(~d\Tb~\t\tau_J~\G_\l\G^M {\cal S}_1\G^\l~\Th_S~+~
d\Tb~\t\tau_J\gg~\G_\l\G^M {\cal C}_1\G^\l~\Th_C~)
\nn\\
&-&(~d\Tb~\G^\l\G^M \G^n~\t\tau_J~d\T)(
\Thb_S~\G_n{\cal S}_1\G_\l~\Th_S~+~\Thb_C~\G_n{\cal S}_1\G_\l~\gg~\Th_C~+~
\nn\\
&&~~~~~~~~~~~~~~~~~~~~~~~+~\Thb_C~\G_n{\cal C}_1\G_\l~\gg~\Th_S~
-~\Thb_S~\G_n{\cal C}_1\G_\l~\gg~\Th_C~)
\nn\\
&+&(~d\Tb~\G^\l\G^M \G^n~\t\tau_J\gg~d\T)(
\Thb_S~\G_n{\cal S}_1\G_\l~\gg~\Th_S~+~\Thb_C~\G_n{\cal S}_1\G_\l~\Th_C
\nn\\
&&~~~~~~~~~~~~~~~~~~~~~~~+~\Thb_C~\G_n{\cal C}_1\G_\l~\Th_S~
-~\Thb_S~\G_n{\cal C}_1\G_\l~\Th_C~).
\label{U}
\eea
\medskip

These $Z$'s are surface terms. It is seen from the fact that \bref{QQQ2}
is a closed form
\be
d~(~\D_{[\ep'}~q^2_{\ep]}~)~=~0.
\ee
The boundary conditions of the D-branes are chosen so that 
there is no supercurrent flows from the boundaries
or they are periodic for the compact configurations. 
The surface terms give topological charges,
which may appear from the non-trivial configurations of 
the coordinates $X^m$ and the gauge fields $A_\mu$.
Assuming the suitable boundary conditions for $\T$, $\Pi$
and $\CF$ the possible topological terms arise only from the 
$\ba\ep {\cal S}_1\ep$ term in \bref{QQQ2}
and $U^M_J$ in the SUSY algebra \bref{U} becomes
\bea
U^M_J&=&\frac12~Tr~(~\t\tau_J~\G^M~{\cal S}_B(dX)~\tau_1~)~e^{dA}~.
\label{Utop}
\eea
where the definition of ${\cal S}_B(dX)$ is given from \bref{defS}. 
The expression of the topological central charges $Z_1$ and $Z_2$ 
of \bref{qqz}, up to the Gauss law constraint, are listed in table
\ref{tblz}.
\vskip 6mm

\begin{table}[hbtp]
\caption[SUSYcc]{SUSY central charges, $Z_1$ and $Z_2$}\label{tblz}
\begin{center}
\begin{tabular}{|c|c|c|c|}
\hline
 & $Z_1^{(1)}$ & $Z_1^{(5)}$ & $Z_2^{(3)}/i$ \\
\hline
D-1&$dX$&$0$&$0$\\
\hline
D-3&$dXdA$&$0$&$(dX)^3$\\
\hline
D-5&$dX(dA)^2/2!$&
$((dX)^5+^*(dX)^5)/2$&
$(dX)^3dA$\\
\hline
D-7&$dX(dA)^3/3!$&
$((dX)^5+^*(dX)^5)/2(dA)$&
$(dX)^3(dA)^2/2!+^*(dX)^7$\\
\hline
D-9
&$dX(dA)^4/4!+^*(dX)^9$
&$((dX)^5+^*(dX)^5)(dA)^2/4$&
$(dX)^3(dA)^3/3!+^*(dX)^7dA$\\
\hline
\end{tabular}
\end{center}
\end{table}
In table 1, for example in $Z^{(3)}_2$, $(dX)^3$ is a symbolic notation of 
$dX^{m_1}dX^{m_2}dX^{m_3}$ and ${}^*(dX)^7$ is 
$\epsilon^{m_1m_2m_3\l_1\l_2\l_3\l_4\l_5\l_6\l_7}
dX_{\l_1}dX_{\l_2}dX_{\l_3}dX_{\l_4}dX_{\l_5}dX_{\l_6}dX_{\l_7}/7!.
$


\section{Summary and Discussions}
\indent

In this paper we develop the canonical formulation of 
general IIB D-p-branes as the first step towards the 
canonical quantization.
In order to perform it we first found the explicit form of WZ actions for IIB
D-branes. The differential equation determining the general WZ action 
given in \bref{wzcond} is integrated by utilizing differential forms.
We found convenient spinor forms $\Th_C$ and $\Th_S$ satisfying
\bref{dthcs}. In terms of $\Th_C$ and $\Th_S$, 
in addition to ${\cal S}_B$ and ${\cal C}_B$ ,
 we can handle formulas for D-p-branes 
in the differential forms. 
Especially the WZ action is expressed in the simple form as in \bref{solwz}. 
We determined the set of constraint equations completely in the Hamiltonian
formalism. The algebra of the fermionic constraints tells that a 
half of them are first class constraints generating the kappa
symmetry and another half are second class constraints. 
\medskip

We also constructed SUSY charges and computed Poisson bracket algebra.
We found expressions of SUSY central charges in terms of 
worldvolume variables.
The appearance of the topological charges in the SUSY algebra of extended
models have been discussed in \cite{AGIT}\cite{AIT}\cite{BrgS}.
Recently  Hammer examined the SUSY algebra of IIA D-p-branes \cite{Hammer}. 
The possible forms of the 
conserved SUSY charges and their algebra 
have been discussed mainly
from the point of view of the algebraic consistency. 
Although the full WZ action is not taken into account
topological charges are found using the leading terms.
They correspond, in our discussions of 
IIB theory, to $\T$ independent part of $U^1$ term \bref{Utop}.
The possible interpretations of the central charges are also 
presented in them. 
\medskip

The case for type IIA D-branes can be done straightforwardly.
As is suggested \cite{Shgf} the $\tau_3$ matrix is replaced
by $\G_{11}$ in appropriate ways for the IIA models. However we didn't 
go into it since it requires some more delicate notations.
We leave the IIA theories for future publication \cite{MK2a}. 
\medskip

Interesting issue is the covariant quantization of
the kappa invariant theories, the superstrings and super D-branes. 
Kallosh \cite{RK} discussed it for the D0 brane in detail and drew sketches 
for D-p-branes. 
We discussed it for the D1 brane in detail in \cite{MK} and pointed out 
origins of its difficulties.
Towards the covariant quantization of super-D-branes
it requires further investigations.

\vskip 6mm
{\bf Acknowledgements}\par
\medskip\par
The authors would like to thank Joaquim Gomis, Yuji Igarashi and Katsumi Itoh
for helpful discussions.
M.H. is partially supported by the Sasakawa Scientific Research Grant 
from the Japan Science Society.

\appendix
\section{Definitions and useful formulas}

The definition of ${\cal C}_B$ and ${\cal S}_B$ are same as in \cite{Shgf} 
( their $\psi$ is denoted as $\slPi$ )
\bea
{\cal C}_B(\slPi)&=&\sum_{\l=0}~\gg^{\l+1}\frac{\slPi^{2\l}}{(2\l)!}~~~~=~
\gg~+~\frac{\slPi^2}{2!}~+~\gg~\frac{\slPi^4}{4!}~+~...~,
\nn\\
{\cal S}_B(\slPi)&=&\sum_{\l=0}~\gg^{\l}\frac{\slPi^{2\l+1}}{(2\l+1)!}~=~
\slPi~+~\gg~\frac{\slPi^3}{3!}~+~\frac{\slPi^5}{5!}~+~...~.
\eea
\medskip

For any one form vector such as $\Pi^m$ we have 

\bea
\G_m~\slPi^s&=&(-)^s~\slPi^{s}~\G_m~+~2~s~\Pi_m~\slPi^{s-1}
\eea
then
\bea
{\cal S}_B~\G_m&=&-~\G_m~{\cal S}_B~+~2~\gg~{\cal C}_B~\Pi_m~,
\nn\\
{\cal C}_B~\G_m&=&\G_m~{\cal C}_B~+~2~{\cal S}_B~\Pi_m.
\label{SGcom}
\eea

For {\bf odd} variation $\D$ and for any one form vector such as $\Pi^m$
\bea
\D~\slPi^\l&=&\l(\l-1)~(-)^\l~\slPi^{\l-2}~(\Pi_m~\D~\Pi^m)~-~
(-)^\l~\l~\slPi^{\l-1}~(\G_m~\D~\Pi^m).
\label{delpiL}
\eea
Using them we have
\bea
\D~{\cal S}_B&=&\gg~{\cal C}_B~(\G_m~\D~\Pi^m)~-~\gg~{\cal S}_
B~(\Pi_m~\D~\Pi^m),
\nn\\
\D~{\cal C}_B&=&-~{\cal S}_B~(\G_m~\D~\Pi^m)~+~\gg~{\cal C}_B~(\Pi_m~\D~\Pi^m).
\label{delSBCB}\eea
\medskip

The formula for {\bf even} variation $\Delta$ are obtained
by writing $~\D=c\Delta~$ and removing the odd constant $c$
from the both sides.
From \bref{delpiL}
\bea
\Delta~\slPi^\l&=&~-~\l(\l-1)~\slPi^{\l-2}~(\Pi_m~\Delta~\Pi^m)~+~
\l~\slPi^{\l-1}~(\G_m~\Delta~\Pi^m).
\label{DelpiL}
\eea
From \bref{delSBCB} we have
\bea
\Delta~{\cal S}_B&=&\gg~{\cal C}_B~(\G_m~\Delta~\Pi^m)~-~
\gg~{\cal S}_B~(\Pi_m~\Delta~\Pi^m),
\nn\\
\Delta~{\cal C}_B&=&+~{\cal S}_B~(\G_m~\Delta~\Pi^m)~-~
\gg~{\cal C}_B~(\Pi_m~\Delta~\Pi^m)~,
\label{DelSBCB}
\eea
\medskip

For two form $\CF$ 
\bea
\D~e^\CF&=&e^\CF~\D~\CF,~~~~~~~~~~~~~~~
\Delta~e^\CF~=~e^\CF~\Delta~\CF.
\label{DelF}
\eea
Up to now we have used only properties of gamma matrices.
\vskip 6mm

In this paper we consider three kind of derivations,
exterior derivative, global SUSY transformations and kappa transformations;
$~\D~~\rightarrow~~d,~~c\D_\ep,~~c\D_\k~$, where c is an odd constant.
Under these derivations
\bea
d\T,&&~~~~~~~~~~~~~~~~d\Pi^m~=~d\Tb~\G^m~d\T,~~~~~~~~~ 
d\CF~=~d\Tb~\gg~\slPi~d\T,
\\
\nn\\
\D\T&=&c\ep,~~~~~~~~~~~~~\D~\Pi^m~=~0,~~~~~~~~~~~~~~~~~~~\D~\CF~=~0,
\\
\nn\\
\D\T&=&c~\D_\k~\T,~~~~~~~\D\Pi^m~=~2~\D\Tb~\G^m~d\T,~~~~~~~ 
\D\CF~=~2~\D~\Tb~\gg~\slPi~d\T,
\label{delkt}
\eea
respectively. 
We can express them collectively as
\bea
\D_\chi\Pi^m&=&\ba\chi~\G^m~d\T,~~~~~~~~~
\D_\chi\CF~=~\ba\chi~\gg~\slPi~d\T,
\eea
and
\bea
\chi&=&d\T~~~~~~~~~for~~~~~~~~d,~~~~~~~~~
\\
\chi&=&0~~~~~~~~~~~for~~~~~~~~SUSY,
\\
\chi&=&2~\D\T~~~~~~~for~~~~~~~~kappa.
\label{delkchi}\eea

Using the notation \bref{BBB}, ${\cal S}_1$ and ${\cal C}_1$ in 
\bref{defs1c1} satisfy 
\bea
\D_\chi~{\cal S}_1&=&\frac12~\gg~[~\slB_{\chi,d\T}~{\cal C}_1~
+~{\cal C}_1~\slB_{\chi,d\T}~],
\\
\D_\chi~{\cal C}_1&=&~\frac12~~[~\slB_{\chi,d\T}~{\cal S}_1~-~{\cal S}_1~
\slB_{\chi,d\T}~].
\label{dchiCS1}
\eea
Especially for $\D=d$ it gives \bref{dCS}.
\vskip 6mm

Another important quantities we use are
j-form spinor $\Theta_j$ (with parity j+1) defined by
\bea
\Th_{j}&=&\slB~\Th_{j-1}~=...=~\slB^j~\Th_0,~~~~~~~~~~\Th_{0}~\equiv~\T.
\eea
It follows
\bea
\Thb_{j}&=&(-)^{j-1}~\Thb_{j-1}~\slB~=...=~(-)^{\frac{j(j-1)}{2}}\Tb~\slB^j.
\eea
\medskip

We further define
\bea
\Th_C&=&\sum_{n=0}~\frac{1}{(4n+1)!!}~(-\gg)^n~\Th_{2n}~=~
\sum_{n=0}~\frac{1}{(4n+1)!!}~(-\gg)^n~\slB^{2n}~\T,
\nn\\
\Th_S&=&\sum_{n=0}~\frac{1}{(4n+3)!!}(-\gg)^{n+1}~\Th_{2n+1}~=~
\sum_{n=0}~\frac{1}{(4n+3)!!}(-\gg)^{n+1}~\slB^{2n+1}\T,
\eea
or equivalently
\bea
\Thb_C&=&\sum_{n=0}~\frac{1}{(4n+1)!!}~\Thb_{2n}~(-\gg)^n~=~
\sum_{n=0}~\frac{1}{(4n+1)!!}~\Tb~\slB^{2n}~\gg^n,
\nn\\
\Thb_S&=&\sum_{n=0}\frac{1}{(4n+3)!!}\Thb_{2n+1}(-\gg)^{n+1}~=~
-~\sum_{n=0}\frac{1}{(4n+3)!!}\Tb~\slB^{2n+1}\gg^{n+1}.
\eea

We start with finding  formulas of variations of $\Th$'s under
general odd derivations of $\T$. 
\bea
\D~\Thb_j~=&-&\frac{2j+1}{2}~\Thb_{j-1}~\D~\slB
\nn\\
&+&(-)^j(j+1)(j-1)~\Thb_{j-2}~(\D B~B)~-~
(-)^{\frac{j(j-1)}{2}}~\ba\rho_\D~\slB^{j-1},
\label{delThj}
\eea
where
\bea
\ba\rho_\D~\equiv~\frac12~\Tb~d\slB_{\T,\D\T}~+~d\Tb~\slB_{\T,\D\T}~=~
-\frac12~\Tb~(\D\slB)~-~(\D\Tb)~\slB,~
\label{defBdel}
\eea
\bref{delThj} is valid also for j=1 since the second term vanish
and is proved by the mathematical induction.

Using it we have 
\bea
\D\Thb_C&=&\D\Tb~-~\frac12~\Thb_S~\D~\slB
\nn\\
&+&\sum_{n=1}\frac{4n^2-1}{(4n+1)!!}\Thb_{2n-2}(-\gg)^n(\D V~V)-
\sum_{n=1}\frac{1}{(4n+1)!!}\ba\rho_\D~\slB^{2n-1}\gg^n
\nn\eea
and
\bea
\D\Thb_S&=&\frac12~\Thb_C~\gg~\D~\slB
\nn\\
&-&\sum_{n=1}\frac{2n(2n+2)}{(4n+3)!!}\Thb_{2n-1}(-\gg)^{n+1}(\D V~V)+
\sum_{n=0}\frac{1}{(4n+3)!!}\ba\rho_\D~\slB^{2n}\gg^{n+1}.
\label{delThS}\eea
\medskip

For $~\D=d$,~ \bref{delThj} becomes
\bea
d~\Thb_j&=&-~\frac{2j+1}{2}~\Thb_{j-1}~d\slB,
\nn\\
d~\Th_j&=&~~\frac{2j+1}{2}~d\slB~\Th_{j-1}~,~~~~~~(j=1,2,...)
\label{dTh1}
\eea
since the second and third terms of \bref{delThj} disappear by
\be
(d V~V)~=~0,~~~~~~\rho~=~(\frac12~\Tb~d\slB~+~d\Tb~\slB~)~=~0.
\ee
Similarly \bref{delThS} becomes  for $~\D=d$ ~
\bea
d~\Thb_C&=&d\Tb~-~\frac12~\Thb_S~d\slB,~~~~~~~~~~
d~\Th_C~=~d\T~+~\frac12~d\slB~\Th_S,
\nn\\
d~\Thb_S&=&~~~~\frac12~\Thb_C~\gg~d\slB,~~~~~~~~~~~
d~\Th_S~=~~~~-~\frac12~\gg~d\slB~\Th_C.
\label{dthcs1}
\eea
\medskip

We express \bref{delThS} as
\bea
\D~\Thb_C&=&\D\Tb~+~d~\Sigb_C^1(\D\T)~+~\Sigb_C^2(\D\T)
\nn\\
\D~\Thb_S&=&d~\Sigb_S^1(\D\T)~+~\Sigb_S^2(\D\T)
\label{delThS3}\eea
where $\Sigb(\D\T)$'s are arranged so that they do not contain $~d(\D\T)~$.
The expressions for $\Sigb(\D\T)$'s are
\bea
\Sigb^1_C(\D\T)
&=&\sum_{n=0}\frac{2n+2}{(4n+5)!!}~
[~\Thb_{2n+1}~\slB_{\T,\D\T}~+~(2n+1)~\Thb_{2n}~(V_{\T,\D\T}~V)~](-\gg)^{n+1}
\nn\\
\Sigb^1_S(\D\T)
&=&\sum_{n=0}\frac{2n+1}{(4n+3)!!}~
[~-~\Thb_{2n}~\slB_{\T,\D\T}~+~2n~\Thb_{2n-1}~(V_{\T,\D\T}~V)~](-\gg)^{n+1}
\label{tSig1S}
\eea
and $\Sigb^2$'s are related to $\Sigb^1$'s by
\bea
\Sigb^2_C(\D\T)
&=&
\frac{1}{2}~\Sigb^1_S(\D\T)~d\slB~-~\Thb_S~\slB_{\D\T, d\T},
\\
\Sigb^2_S(\D\T)
&=&
(~-~\frac{1}{2}~\Sigb^1_C(\D\T)~d\slB~+~\Thb_C~\slB_{\D\T, d\T}~)~\gg.
\label{tSig2}
\eea
\vskip 6mm

Corresponding relations for even variation is
from \bref{delThj}
\bea
\Delta~\Thb_j~=&-&(-)^j~\frac{2j+1}{2}~\Thb_{j-1}~\Delta~\slB
\nn\\
&+&(j+1)(j-1)~\Thb_{j-2}~(V~\Delta V)
\nn\\
&-&(-)^{\frac{j(j-1)}{2}}~
\{\frac12~\Tb~\Delta \slB~-~\Delta\Tb~\slB~\}~
\slB^{j-1}
,~~~~~~~~~~~j~\geq~1.
\eea
where
\bea
\Delta~V
&=&V_{\Delta\T,d\T}~+~V_{\T,\Delta d\T}~=~
-~d~V_{\T,\Delta\T}~+~2~V_{\Delta\T,d\T}.
\eea
From \bref{delThS}
\bea
\Delta~\Thb_C&=&\Delta\Tb~-~\frac{1}{2}~\Thb_{S}~\Delta~\slB
\nn\\
&+&\sum_{n=1}\frac{(2n+1)(2n-1)}{(4n+1)!!}~\Thb_{2n-2}~(-\gg)^n~(V~\Delta V)
\nn\\
&-&\sum_{n=1}\frac{1}{(4n+1)!!}~
\{\frac12~\Tb~\Delta \slB~-~\Delta\Tb~\slB~\}
\slB^{2n-1}~\gg^n,
\label{DelTC}
\eea
\bea
\Delta~\Thb_S&=&-~\frac{1}{2}~\Thb_{C}~\gg~\Delta~\slB
\nn\\
&+&\sum_{n=1}\frac{(2n+2)2n}{(4n+3)!!}~\Thb_{2n-1}~(-\gg)^{n+1}~(V~\Delta V)
\nn\\
&+&\sum_{n=0}\frac{1}{(4n+3)!!}~
\{\frac12~\Tb~\Delta \slB~-~\Delta\Tb~\slB~\}
\slB^{2n}~\gg^{n+1}.
\label{DelTS}
\eea
Corresponding to \bref{delThS3} we write them as
\bea
\Delta~\Thb_C&=&\Delta\Tb~+~d~\Sigb_C^1[\Delta\T]~+~\Sigb_C^2[\Delta\T],
\nn\\
\Delta~\Thb_S&=&d~\Sigb_S^1[\Delta\T]~+~\Sigb_S^2[\Delta\T],
\label{delThS4}\eea
where $\Sigb^1[\Delta\T]$ is obtained from $\Sigb^1(\D\T)$ by replacing
$\D\T$ by $c\Delta\T$ and remove $c$ from the left and put minus on it (
due to the existence of $d$ on it in  the definition).
\bea
\Sigb^1_C[\Delta\T]&=&\sum_{n=0}\frac{2n+2}{(4n+5)!!}~
[~\Thb_{2n+1}~\slB_{\T,\Delta\T}~-~(2n+1)~\Thb_{2n}~(V_{\T,\Delta\T}~V)~]
(-\gg)^{n+1},
\nn\\
\Sigb^1_S[\Delta\T]&=&\sum_{n=0}\frac{2n+1}{(4n+3)!!}~
[~~\Thb_{2n}~\slB_{\T,\Delta\T}~+~2n~\Thb_{2n-1}~(V_{\T,\Delta\T}~V)~]
(-\gg)^{n+1}.
\label{tSig1SD}
\eea
\medskip
$\Sigb^2[\Delta\T]$ is obtained from $\Sigb^2(\D\T)$ 
and satisfies, from \bref{tSig2}
\bea
\Sigb^2_C[\Delta\T]&=&
-~\frac{1}{2}~\Sigb^1_S[\Delta\T]~d\slB~-~\Thb_S~\slB_{\Delta\T,d\T},
\nn\\
\Sigb^2_S[\Delta\T]&=&
(~\frac{1}{2}~\Sigb^1_C[\Delta\T]~d\slB~-~\Thb_C~\slB_{\Delta\T,d\T}~)~\gg.
\label{tSig2D1}
\eea
\vskip 6mm

The variations of $\Sigma^1[g]$'s under even variation $~\Delta_f\T=f~$ is
expressed as
\bea
\Delta_{[f}(\Sigb^1_C[g_{]}])~\equiv~d\Lambda^1_C~+~\Lambda^2_C~,
\nn\\
\Delta_{[f}(\Sigb^1_S[g_{]}])~\equiv~d\Lambda^1_S~+~\Lambda^2_S~
\label{delfSig1}
\eea
so that $\Lambda$'s do not contain $df$ and $dg$. 
The explicit form of $\Lambda$'s are
\bea
\Lambda^1_C&=&\sum_{n=0}\frac{(2n+2)(2n+1)}{(4n+5)!!}~
[~\frac12\Thb_{2n}\slB_{\T,[f}\slB_{\T,g]}~-~2n~\Thb_{2n-1}
\slB_{\T,[f}(V~V_{\T,g]})
\nn\\
&&+~2n(2n-1)~\Thb_{2n-2}(V~V_{\T,f})(V~V_{\T,g})~]~(-\gg)^{n+1},
\nn\\
\Lambda^1_S&=&\sum_{n=1}\frac{(2n+1)(2n)}{(4n+3)!!}~
[~\frac12\Thb_{2n-1}\slB_{\T,[f}\slB_{\T,g]}~+~(2n-1)~\Thb_{2n-2}
\slB_{\T,[f}(V~V_{\T,g]})
\nn\\
&&+~(2n-1)(2n-2)~\Thb_{2n-3}(V~V_{\T,f})(V~V_{\T,g})~]~(-\gg)^{n+1}.
\label{Lam1}
\eea
$\Lambda^2$'s are related to $\Lambda^1$'s by
\bea
\Lambda^2_S&=&-~\frac12\Lambda^1_C\gg d\slB~-~\Thb_C\gg\slB_{f,g}~-~
\Sigb^1_C[{}_{[}g]\gg\slB_{f],d\T},
\nn\\
\Lambda^2_C&=&\frac12\Lambda^1_S d\slB~+~\Thb_S\slB_{f,g}~-~
\Sigb^1_S[{}_{[}g]\slB_{f],d\T}.
\label{lam22}
\eea
\vskip 6mm
\eject

Finally we give lists of "signs" when spinors are interchanged in the
bi-linear forms in table.\ref{U}.

\vskip 6mm
\begin{table}[hbtp]
\caption[signs]{Signs of bi-spinor forms under transposition} \label{U1}
\begin{center}
\begin{tabular}{|l|c|c|c|c|}
\hline
 & \multicolumn{4}{c|}{$(sign)$ ~~~for $(\chi,\phi)$ with their sign}\\
\cline{2-5}
bi-linear forms&(odd,odd)&(odd,even)&
(even,odd)&(even,even)\\
\hline
&&&&\\
 $\ba\chi~{\cal S}_1~\phi~=~(sign)~\ba\phi~{\cal S}_1~\chi$ &
$-$&$-$&$-$&+\\
\hline
&&&&\\
$\ba\chi~{\cal C}_1~\phi~=~(sign)~\ba\phi~{\cal C}_1~\chi$ &
$-$&+&+&+\\
\hline
for odd $V$&
\multicolumn{4}{c|}{~~~}\\
\hline
&&&&\\
$\ba\chi~{\cal S}_1~\slB~\phi~=~(sign)~\ba\phi~\slB~{\cal S}_1~\chi$
&$-$&+&+&+\\
\hline
&&&&\\
$\ba\chi~{\cal C}_1~\slB~\phi~=~(sign)~\ba\phi~\slB~{\cal C}_1~\chi$
&+&+&+&$-$\\
\hline
for even $V$&\multicolumn{4}{c|}{~~~}\\
\hline
&&&&\\
$\ba\chi~{\cal S}_1~\slB~\phi~=~(sign)~\ba\phi~\slB~{\cal S}_1~\chi$
&+&+&+&$-$\\
\hline
&&&&\\
$\ba\chi~{\cal C}_1~\slB~\phi~=~(sign)~\ba\phi~\slB~{\cal C}_1~\chi $
&+&$-$&$-$&$-$\\
\hline
\end{tabular}\end{center}
\end{table}
\eject


\end{document}